\documentclass[twocolumn,floats,floatfix,aps,pra]{revtex4-2}
\usepackage{graphicx} 

\usepackage[T1,T2A]{fontenc}
\usepackage[utf8]{inputenc}

\usepackage{amsfonts,amssymb,amsmath}
\usepackage{physics}
\usepackage{color,calc}
\usepackage{bm}
\usepackage{amsbsy,amsmath}
\usepackage{array}
\usepackage{booktabs}
\usepackage{graphicx}
\usepackage{graphics}
\usepackage{eepic,epsfig}
\usepackage{physics}

\usepackage{soul}

\usepackage{microtype}

\usepackage[breaklinks,hidelinks,colorlinks=true,linkcolor=blue,citecolor=blue,filecolor=blue,urlcolor=blue]{hyperref}

\usepackage[dvipsnames]{xcolor}
\definecolor{myblue}{RGB}{0, 0, 255}
\usepackage[toc,page]{appendix}

\def\be{ \begin{equation} }
\def\ee{ \end{equation} }
\def\bea{ \begin{eqnarray} }
\def\eea{ \end{eqnarray} }
\def\bse{ \begin{subequations} }
\def\ese{ \end{subequations} }
\def\ba{ \begin{array} }
\def\ea{ \end{array} }
\def\bt{ \begin{tabular} }
\def\et{ \end{tabular} }

\long\def\/*#1*/{}

\begin{document}


\title{Line shape of soft photon radiation generated at zero angle in an undulator with a dispersive medium}

\author{Hayk L. Gevorgyan\textsuperscript{\hyperref[1]{1},\hyperref[2]{2}}}
\email{hayk.gevorgyan@aanl.am}
\author{Koryun L. Gevorgyan\textsuperscript{\hyperref[1]{1}}}
\author{Anahit H. Shamamian\textsuperscript{\hyperref[3]{3}}}
\author{Lekdar A. Gevorgian\textsuperscript{\hyperref[4]{4}}}
\email{lekdar@yerphi.am}
\affiliation{
\phantomsection\label{1}{\textsuperscript{1}Experimental Physics Division, A.I. Alikhanyan National Science Laboratory (Yerevan Physics Institute), 2 Alikhanyan Brothers St., 0036 Yerevan, Armenia}\\
\phantomsection\label{2}{\textsuperscript{2}Quantum Technologies Division, A.I. Alikhanyan National Science Laboratory (Yerevan Physics Institute), 2 Alikhanyan Brothers St., 0036 Yerevan, Armenia}\\
\phantomsection\label{3}{\textsuperscript{3}Department of Exact Subjects, Military Academy after Vazgen Sargsyan MoD RA, 155 Davit Bek St., 0090 Yerevan, Armenia}\\
\phantomsection\label{4}{\textsuperscript{4}Matinyan Center for Theoretical Physics, A.I. Alikhanyan National Science Laboratory (Yerevan Physics Institute), 2 Alikhanyan Brothers St., 0036 Yerevan, Armenia}}

\date{\today }

\begin{abstract}
The problem of undulator radiation from a bunch of charged particles, taking into account a medium polarization, is considered. In a dispersive medium, at a zero angle, in addition to hard photons, soft photons are also generated. If the wavelength of the soft photons is greater than or equal to the longitudinal size of the microbunches formed during the FEL process, the microbunches radiate coherently. Consequently, the radiation of the bunch will be partially coherent. As a result, intense, quasi-monochromatic, and directed X-ray photon beams are produced, which can have wide practical applications.
\end{abstract}

\maketitle



\section{Introduction}
In 1947, V. L. Ginzburg proposed the generation of electromagnetic radiation in the submillimeter range using a beam of relativistic electrons in periodic magnetic structures (undulators) \cite{ginzburg1947}. In the theoretical work by Motz \cite{motz1951}, the characteristics of undulator radiation are studied in detail. The first undulator was created by Motz, and experimental studies of the characteristics of undulator radiation were conducted using a beam of relativistic electrons from a Stanford linear accelerator \cite{motz1953}. It should be noted that Vavilov-Cherenkov radiation was also generated in the experiment.

At Vavilov's initiative, Cherenkov observed the blue glow of liquids under the influence of $\gamma$-rays in the conducted experiment \cite{Cherenkov1934}. Vavilov asserted that the observed radiation, which had also been noted in earlier works by other scientists, was not luminescence but a new phenomenon caused by the radiation of fast electrons generated during the irradiation of matter \cite{Vavilov1934}. In the work of Tamm and Frank \cite{Tamm1937}, the mechanism of Vavilov-Cherenkov (V.C.) radiation was identified. The established quantitative classical theory was later confirmed by quantum considerations (Ginzburg, 1940). V.C. radiation detectors are used to measure the velocities of relativistic charged particles, as mass spectrometers, and as discriminators.

It was proposed to use optical radiation generated in an undulator containing a transparent medium to measure the energy of a relativistic electron beam. Following the publication of works \cite{godvin1969, korkhmazian1970}, a large number of theoretical and experimental studies have been dedicated to investigating the characteristics of undulator radiation in a hard photon frequency range. 

The undulator radiation in a dispersive medium was investigated in \cite{gevorgian19771979}. In such a medium, there is an energy threshold for the formation of radiation. When the energy of a charged particle significantly exceeds the threshold energy, two photons are emitted at a zero angle: a soft photon with energy that does not depend on the particle's energy, and a hard photon with energy that depends quadratically on the particle's energy. As the particle energy approaches the threshold energy, the particle spectrum narrows, i.e., a charged particle with threshold energy radiates photons with a specific energy at a zero angle. The spectrum narrowing effect has significant practical importance. In a dispersive medium, at particle energies much higher than the threshold energy, the frequency of the measured soft photons almost does not depend on the energy of the charged particle beam and constitutes half the frequency of the photons formed during spectrum narrowing. Consequently, using undulators filled with a dispersive medium can enhance both the efficiency of a free-electron laser (FEL) and achieve a high acceleration rate, i.e., an inverse free electron laser (IFEL) \cite{Gevorgian19901991,Gevorgian1993,Gevorgian1994}.

In the work \cite{ginzburg1972}, it was proposed to use optical undulator radiation generated by a bunch of relativistic electrons in a transparent medium to determine the energy of the bunch. In the study \cite{gevorgian1979}, it was shown that optical undulator radiation, like transition radiation, is inseparable from Vavilov-Cherenkov radiation, and its intensity is negligibly small. Therefore, as noted in the section \textit{On an unsuccessful attempt to invent a particle counter} in the book by Ginzburg \cite{ginzburg1979}, an undulator counter in a medium will not work, and the mistake made was in considering these radiations independently of each other.

The optical radiation produced by bombarding the surfaces of massive targets of various metals with a beam of non-relativistic electrons was investigated in many experimental studies before the discovery \cite{Ginzburg1946} of transition radiation. 

In the early works \cite{Frank1965,Ter-Mikaelian1972}, The glow was attributed to luminescence and bremsstrahlung. Following the discovery of transition radiation, it was found that a certain portion of the glow is due to transition radiation. In subsequent experiments, transition radiation was studied when electrons passed through metallic targets and films. A comparison of the results of experiments with the theory of transition radiation was conducted in \cite{Silin1964}. In the experiments, the glow brightness increases proportionally with the energy of the electrons; and the presence of a peak, in the transparency region of metals, agrees with the theory of transition radiation. However, the presence of an unpolarized component of light remains unexplained.

The cause of the abnormally high intensity of transition radiation observed in experiments during the oblique incidence of an electron beam on the interface between two media \cite{Blanckenhagen1964,Boersch1965} was also not identified. During grazing incidence, the radiation intensity exceeds that of transition and bremsstrahlung radiation by 1 to 2 orders of magnitude, and the position of the maximum in the radiation spectrum does not coincide with the transparency band of the material \cite{Ter-Mikaelian1972}. In \cite{Gevorgian1979a}, it was shown that the reason for the discrepancies between the experimental data and the characteristics of transition radiation is the irregularly periodic motion of electrons above the rough surface of the target, caused by the accumulation of free electrons at the peaks of irregularly spaced surface roughnesses. This explains the presence of an unpolarized light component both during normal electron incidence and the abnormally high intensity during grazing incidence. The frequency distribution of the radiation obtained in \cite{Gevorgianian1979b} is in good agreement with the experimental data. The abnormally high intensity is attributed to undulator radiation. Our study is dedicated to determining the line shape of radiation at a zero angle for both hard and soft photons, generated in the undulator with a dispersive medium. The expression obtained for the number of soft photons can be used in the process of coherent radiation of microbunches.

In the work \cite{Garibyan1959}, the energy losses of a relativistic charged particle moving uniformly across the boundary between two media were studied and found to be proportional to its energy. In \cite{Barsukov1959}, it was confirmed that these losses are due to X-ray transition radiation (XTR). The radiator of an XTR detector is a medium with random inhomogeneities. Following Godvin's note on the possibility of using optical undulator radiation (OUR) for spectroscopy and especially Korkhmazian's work on X-ray optical undulator radiation (XOUR), numerous theoretical studies emerged. The first experiment on XOUR was conducted on the ARUS electron synchrotron \cite{Alikhanyan1972}. This work also proposed the idea of generating $\gamma$-quanta through undulator radiation formed in a multidomain structure of a magnetic film. In studies \cite{Bratman1982, Genkin1982}, OUR generated by a magnetic film and by grazing passage of relativistic electrons over the surface of a ferromagnet was investigated. Based on the formulas obtained by the authors, it was claimed that the radiation intensity is on the order of or greater than the intensity of radiation produced by the channeling of charged particles in crystals. This problem was first studied by Kumakhov \cite{Kumakhov1976}. However, as shown in \cite{Gevorgian1983a}, the formulas obtained in the aforementioned works are incorrect. The actual radiation intensities are six orders of magnitude lower. As a result, preparations for experiments in many scientific centers were halted. 

Madey studied the process of stimulated undulator radiation \cite{Madey1971}, known as the Free Electron Laser (FEL), and demonstrated that the radiation intensity gain is proportional to the derivative of the spontaneous emission line shape. The stimulated process was postulated by Einstein in 1916 to explain thermal equilibrium in a system of many particles. The existence of this phenomenon has been proven within the framework of classical electrodynamics. In quantum electrodynamics, it has a more consistent interpretation and has been experimentally confirmed. In the 1933 article by Kapitza and Dirac \cite{Kapitza1933}, stimulated Compton scattering was considered. The interaction of photons emitted at a zero angle with the electrons of the bunch during the FEL process leads to the grouping of electrons into microbunches with a size equal to the radiation wavelength. Consequently, the radiation of the bunch is partially coherent, which occurs in the process of X-ray FELs. The work \cite{Gevorgyan1982} examined the case where FEL amplification is due to the effect of partially coherent radiation from a bunch with a longitudinal asymmetric distribution. In \cite{Cocke1979}, the possibility of optical FEL amplification in a gaseous medium in the absence of Vavilov-Cherenkov radiation was considered. In the experiment \cite{Pantell1986}, the opposite phenomenon was observed. This result is related to a violation of the synchronism condition: a shift in the spectrum of spontaneous emission during the FEL process. The synchronism condition is maintained if the gas density along the undulator changes according to a specific law \cite{Gevorgian1994}. This method allows for an increase in the acceleration rate of charged particles \cite{Gevorgian1993,gevorgian2005,widemann2001}.

The characteristics of undulator radiation formed in a dispersive medium were investigated in the study \cite{gevorgian19771979}. If the energy of the charged particle bunch significantly exceeds the threshold energy for radiation formation, the frequency of soft photons becomes almost independent of the energy. This property can be utilized both to enhance the efficiency of FEL and to achieve a high acceleration rate \cite{gevorgian2005}.

Our study is dedicated to determining the line shape of radiation at a zero angle for both hard and soft photons, generated in the undulator with a dispersive medium. The expression obtained for the number of soft photons can be used in the process of coherent radiation of microbunches.

In Sec.~\ref{Sec:2}, the process of radiation formation during the interaction of a charged particle with an external field and/or a medium is considered. Sec.~\ref{Sec:3} investigates the spectral distribution of undulator radiation (UR) generated in a dispersive medium. In Sec.~\ref{Sec:4}, an expression is derived for the number of soft (hard) photons produced at zero angle by an ultrarelativistic charged particle in an undulator with a medium. Finally, Sec.~\ref{Sec:concl} presents the conclusions.

\section{On Representing the Spectral Distribution of Radiation of Any Nature as an Integral over the Trajectory of a Charged Particle}\label{Sec:2}

The system of four interrelated Maxwell equations of the electromagnetic field forms the foundation of electrodynamics. The microscopic form of these first-order partial differential equations, when using scalar and vector potentials, is reduced to two second-order equations. By utilizing the freedom in defining the potentials, it is possible to choose a system of potentials that satisfies the Lorenz condition. In this case, the potentials satisfy individual inhomogeneous wave equations, which have the same structure. If the gauge scalar function satisfies the homogeneous wave equation, then the Lorenz condition remains valid for the potentials satisfying this condition before transformation. The Lorenz gauge is independent of the choice of the coordinate system. A particular solution of the inhomogeneous wave equation for scalar and vector potentials, for a confined distribution of charges and currents in the absence of boundaries, is determined through the retarded Green's function, i.e., in accordance with the causality condition. The electromagnetic field of a point charge, determined by the Liénard–Wiechert potentials, includes not only the static field but also the transverse radiation field, which depends on acceleration. The total energy radiated per unit solid angle is determined by integrating the power over time, which is expressed in terms of the electric field strength of the radiation. The Fourier amplitude of the radiation field is expressed as an integral along the charge trajectory. It is assumed that the observation point is sufficiently distant from the region of space where the radiation field is formed, such that it is observed under a small solid angle. Integration is carried out over the time interval during which the acceleration of the charged particle is non-zero. In the frequency-angular distribution of the radiation energy, obtained based on the microscopic form of the Maxwell equations, by replacing the speed of light in a vacuum $c$ with $c/\sqrt{\epsilon (\omega)}$ and the charge $e$ with $e/\sqrt{\epsilon (\omega)}$, where $\omega$ is the radiation frequency and is much greater than a plasma frequency $\omega_p$; we obtain the energy distribution of the radiation produced in the dispersive medium  with the dielectric permittivity
\be
\epsilon(\omega) = 1 - \frac{\omega_p^2}{\omega^2}.
\ee

Dividing the energy distribution of the radiation by $\hbar \omega$, where $\hbar = h/(2\pi)$ and $h$ is Planck's constant, yields the following frequency-angular distribution for the number of radiated photons \cite{Jackson1999}

\be
\frac{d^3 N}{d\omega d(\cos{\vartheta}) d\varphi} = \frac{\alpha \omega \sqrt{\epsilon (\omega)}}{4\pi^2} \lvert \vb*{I}(\omega, 
\vartheta) \rvert^2,
\ee
where  $d(\cos{\vartheta}) d\varphi$ represents the solid angle of radiation: the polar angle $\vartheta$ is measured from the unit vector in the direction of radiation $\vb*{n} = - \sin{\vartheta} \cos{\varphi} \, \vb*{i} - \sin{\vartheta} \sin{\varphi} \, \vb*{j} + \cos{\vartheta} \, \vb*{k}$, and $\varphi$ is the azimuthal angle. The constant $\alpha = 1 / 137$ is the fine-structure constant. The electromagnetic field of the radiation is defined by the following time integral:
\be\label{Iomegavartheta}
\begin{aligned}
\vb*{I} (\omega, \vartheta) &= \int \vb*{a} (t) \exp{i \omega \left(t - \frac{\sqrt{\epsilon (\omega)}}{c} \vb*{n} \cdot \vb*{r}(t) \right)} dt, \\
\vb*{a} (t) &= [\vb*{n} \times [\vb*{n} \times \vb*{\beta}]],
\end{aligned}
\ee
where $\vb*{\beta} (t)$ is the particle's velocity in units of $c$, and $\vb*{r} (t)$ is the particle's trajectory. The integration in \eqref{Iomegavartheta} is performed only over the time interval during which the charged particle interacts with the medium and/or external fields.

When the velocity of the charged particle is constant, radiation occurs when the argument of the exponent goes to zero. This condition corresponds to the law of conservation of energy-momentum and relates the photon energy to the direction of radiation. It is satisfied in the case of Vavilov-Čerenkov radiation $\beta \sqrt{\epsilon (\omega)} > 1$, as well as in transition radiation due to the difference $\Delta\epsilon (\omega)$ at the boundary of two media. When the velocity of the charged particle changes, the law of conservation of energy-momentum holds, taking into account the integrand factor $\vb*{a}(t)$.

\section{The Line Shape of Soft and Hard Photons Generated in a Dispersive Medium of a Planar Sinusoidal Magnetic Undulator}\label{Sec:3}

To the extent that small energy losses due to radiation justify a classical approach to the problem, the energy of the particle is conserved. As the transverse component of the velocity of a particle with energy $\gamma m c^2$ changes; where $\gamma$ is the Lorentz factor and $mc^2$ is the rest energy; the longitudinal component also changes. However, radiation caused by changes in the longitudinal component of velocity is smaller by a factor of $\gamma^2$ than the radiation caused by changes in the transverse component. Therefore, we will assume that the particle moves in the longitudinal direction of the undulator with a root-mean-square velocity $\beta_\parallel = \sqrt{\langle\beta_\parallel (t)^2\rangle}$. Then, due to energy conservation, $\beta^2 = \langle\beta_\perp (t)^2\rangle + \langle\beta_\parallel (t)^2\rangle = \langle\beta_\perp (t)^2\rangle + \beta_\parallel^2$, it follows that the energy of the longitudinal motion of the particle $\gamma_\parallel = \frac{1}{\sqrt{1-\beta_\parallel^2}}$ is less than the total energy $\gamma$ by a factor of $\sqrt{Q} = \sqrt{1+\gamma^2 \langle\beta_\perp (t)^2\rangle}$. The transverse velocity of the particle in the magnetic field of a planar undulator (a one-dimensional oscillator) varies according to the law $\vb*{\beta}_\perp (t) = - \beta_\perp \sin{(\Omega t)} \, \vb*{j}$, where $\beta_\perp$ is the undulator parameter, and $\Omega$ is the oscillation frequency. Thus, the trajectory of the particle is $\vb*{r} (t) = \left(\beta_\perp c/\Omega\right) \cos{(\Omega t)} \, \vb*{j} + \beta_\parallel c \, \vb*{k}$. Note that for a one-dimensional oscillator, $Q = 1+ q^2/2$, where $q = \beta_\perp \gamma$ is the radiation parameter, unlike a two-dimensional oscillator, for which $Q = 1+q^2$.

If the particle in the planar undulator undergoes $n_0$ oscillations with frequency $\Omega$, it is convenient to integrate \eqref{Iomegavartheta} with respect to the variable $\tau = \Omega t$ over the range $[-\pi n_0, \pi n_0]$. Since we are interested in the line shape of the frequency distribution of radiation at $\vartheta \rightarrow 0$, it is easy to show that $\vb*{a} (t) = \vb*{n} \left( - \vartheta \beta_\perp \cos{\varphi} \sin{\tau} + \beta_\parallel \cos{\vartheta}\right) - \vb*{\beta}(t) \approx \vb*{\beta}_\perp (t) = \frac{\vb*{j} \beta_\perp}{2 i} \left(e^{-i\tau} - e^{i\tau}\right)$, and $\frac{\sqrt{\epsilon (\omega)}}{c} \vb*{n} \cdot \vb*{r}(t)$ approaches $\beta_\parallel \sqrt{\epsilon (\omega)} \cos{\vartheta}$.

Using the dimensionless frequency $\xi = \omega/(\Omega \gamma_\parallel^2)$, for the frequency-angular distribution of the number of emitted photons, we obtain:
\bse
\begin{align}
\frac{d^3 N}{d\omega d(\cos{\vartheta}) d\varphi} &= \frac{\alpha (\Omega \gamma_\parallel^2)^2 \xi \sqrt{\epsilon (\omega)} }{\pi^2} \lvert \vb*{I}(\omega, \vartheta) \rvert^2, \\
\vb*{I} = \vb*{j} \frac{\beta_\perp}{2 i \Omega} & \int\limits_{-\pi n_0}^{\pi n_0} \left(e^{-i\tau} -e^{i\tau}\right) e^{i 2 \xi \gamma_\parallel^2 \left(1 - \beta_\perp \sqrt{\epsilon (\xi)} \cos{\vartheta}\right) \tau} d\tau, \label{Ixi}
\end{align}
\ese
where $\sqrt{\epsilon(\xi)} = 1 - \frac{1}{2\gamma_\parallel^2} \left(\frac{r}{2\xi}\right)^2$, $r = \frac{\gamma_\text{th}}{\gamma_\parallel}$, $\gamma_\text{th} = \frac{\omega_p}{\Omega}$.

Considering the expansions $\beta_\parallel \approx 1 - \gamma_\parallel^{-2}/2$, $\cos{\vartheta} \approx 1 - \vartheta^2/2$, the argument of the exponent of the first integrand term, which corresponds to the law of conservation of energy-momentum during radiation, is:
\be
\phi (\xi, \theta) = \frac{1}{\xi} \left((1+\theta^2) \xi^2 - \xi + \frac{r^2}{4} \right),
\ee
where $\theta = \gamma_\parallel \vartheta$. This argument vanishes at the values:
\be
\xi_{\text{1, 2}} = \frac{1}{2(1+\theta^2)} \left(1 \mp \sqrt{1- r^2 (1+ \theta^2)} \right).
\ee

These values are valid (real) if $r \sqrt{1+\theta^2} < 1$. When $r=1$ ($\gamma_\parallel = \gamma_\text{th}$), photons with frequency $\xi_0 = 1$ ($\omega_0 = \Omega \gamma_\text{th}^2 = \omega_p^2/\Omega$) are emitted at zero angle. When $\gamma_\parallel > \gamma_\text{th}$ two photons with frequencies $\omega_1 < \omega_0$ and $\omega_2 > \omega_0$ are emitted at zero angle. For $\gamma_\parallel \gg \gamma_\text{th}$, with accuracy up to small terms of order $r^2$, we have $\xi_1 = r^2/4$ [$\omega = \omega_p^2/(2\Omega)$] and $\xi_2 (\theta) = 1/(1+\theta^2)$ [$\omega_2 (\theta) = \frac{2\Omega \gamma_\parallel^2}{1+\theta^2}$]. Note that for soft photons, the unit of angle $\vartheta$ is 1, while for hard photons ($\theta = \gamma_\parallel \vartheta$): $\gamma_\parallel^{-1}$. Therefore, for soft photons at $\vartheta = 0$, we obtain $\int d(\cos{\vartheta})\, d\varphi = 2\pi$,  and for hard photons: $\int d(\cos{\theta})\,d\varphi = 2\pi$. Considering \eqref{Ixi}, $\vb*{I}(\xi,0) = \vb*{j} \frac{\beta_\perp}{i\Omega} \frac{\sin{\phi(\xi)}}{\phi(\xi)}$. Consequently, $\lvert \vb*{I}\rvert^2 = \frac{\beta_\perp^2}{\Omega^2} F(\xi_{\text{1, 2}})$. The function $F(\xi_{\text{1, 2}}) = \frac{\sin^2 (\pi n_0 \phi(\xi_\text{1, 2}))}{\phi^2 (\xi_\text{1, 2})}$, for $n_0 \gg 1$ is a delta-like function with a peak value of $\pi^2 n_0^2$ and a width of $1/n_0$, regardless of the values of $\xi_\text{1, 2}$. This line shape $F(\xi_{\text{1, 2}})$ of radiation at zero angle is the same for both soft and hard photons.

\section{The Number of Emitted Photons Radiated at Zero Angle}\label{Sec:4}

To an accuracy of order $1/n_0$, the formula 
\be
\begin{gathered}
F(\xi_\text{1, 2}) = \pi^2 n_0 \delta\left(\phi(\xi_\text{1, 2})\right),\\
\phi(\xi) \overset{\Delta}{=} \phi(\xi, 0) = \xi - 1 + \frac{\xi_1}{\xi}, \\
\lvert \phi'(\xi) \rvert_{\xi = \xi_\text{1}} = \frac{1}{\xi_1}, \quad \lvert \phi'(\xi) \rvert_{\xi = \xi_\text{2}} = 1, \\
\delta\left(\phi(\xi_\text{1, 2})\right) = \frac{\delta(\xi- \xi_\text{1, 2})}{\lvert \phi'(\xi) \rvert_{\xi = \xi_\text{1, 2}}} = \begin{cases}
\xi_1 \delta(\xi - \xi_1) & \text{for } \xi_1, \\
\delta(\xi - \xi_2) & \text{for } \xi_2 .
\end{cases},
\end{gathered}
\ee
can be used, where it is considered that $\xi_1 \ll 1< \xi_2 = 1$.

Thus, after integration, for soft and hard photons we have:
\be
\begin{gathered}
N(\xi_1) = N(\omega_1) = 2\pi \alpha n_0 \beta_\perp^2 \gamma_\parallel^4 \xi_1^2 = 2\pi \alpha n_0 \beta_\perp^2 \left(\frac{\omega_p}{2\Omega} \right)^4, \\
N(\xi_2) = N(\omega_2) = 2\pi \alpha n_0 \left(\beta_\perp \gamma_\parallel\right)^2 = 2\pi \alpha n_0 q_\parallel^2, \\
q_\parallel = \beta_\perp \gamma_\parallel = \frac{q}{\sqrt{Q}} \quad q = \beta_\perp \gamma \quad Q = 1 + q^2/2 .
\end{gathered}
\ee

It should be noted that, at a specific energy of the charged particle, the number of soft and hard photons emitted at zero angle is the same.

\section{Conclusion}\label{Sec:concl}
During the formation of microbunches with angstrom-scale length [longitudinal size] and specific energy in a free-electron (positron) laser under axial (planar) channeling of electrons (positrons), coherent emission will occur in the X-ray range at a zero angle. Intense, directed beams of such photons will have significant practical applications.



\end{document}